\documentclass[showpacs,preprintnumbers,amsmath,amssymb,twocolumn]{revtex4}

\usepackage{graphicx}
\usepackage{dcolumn}
\usepackage{txfonts,bm}
\usepackage{hyperref}
\usepackage{color}
\usepackage{epsfig}
\usepackage{multirow}
\usepackage{subfigure}

\begin{document}

\title{Hidden-Charm Tetraquarks and Charged $Z_c$ States}

\author{Lu Zhao$^{1}${\footnote{Email: Luzhao@pku.edu.cn}},
Wei-Zhen Deng$^{1}${\footnote{Email: dwz@pku.edu.cn}}, Shi-Lin
Zhu$^{1,2}${\footnote{Email: zhusl@pku.edu.cn}} }

\affiliation{$^{1}$ Department of Physics and State Key Laboratory
of Nuclear Physics and Technology, Peking University, Beijing
100871, China\\
$^2$Collaborative Innovation Center of Quantum Matter, Beijing
100871, China }

\begin{abstract}

Experimentally several charged axial-vector hidden-charm states were
reported. Within the framework of the color-magnetic interaction, we
have systematically considered the mass spectrum of the hidden-charm
and hidden-bottom tetraquark states. It is impossible to accommodate
all the three charged states $Z_c(3900)$, $Z_c(4025)$ and
$Z_c(4200)$ within the axial vector tetraquark spectrum
simultaneously. Not all these three states are tetraquark
candidates. Moreover, the eigenvector of the chromomagnetic
interaction contains valuable information of the decay pattern of
the tetraquark states. The dominant decay mode of the lowest axial
vector tetraquark state is $J/\psi \pi$ while its $D^*\bar{D}$ and
$\bar{D}^*D^*$ modes are strongly suppressed, which is in contrast
with the fact that the dominant decay mode of $Z_c(3900)$ and
$Z_c(4025)$ is $\bar{D}D^*$ and $\bar{D}^*D^*$ respectively. We
emphasize that all the available experimental information indicates
that $Z_c(4200)$ is a very promising candidate of the lowest axial
vector hidden-charm tetraquark state.

\end{abstract}

 \maketitle{}

\section{INTRODUCTION}\label{introduction}

During the past decade, many charmonium-like states and
bottomonium-like states have been reported by experimental
collaborations such as Belle, $BARBAR$, CDF, D0, LHCb, BESIII and
CLEOc. $X(3872)$ was first observed by Belle Collaboration in the
exclusive decay process $B^{\pm} \rightarrow K^{\pm}\pi^+ \pi^-
J/\psi$ \cite{Choi:2003}. Its mass is very close to the $\bar{D}^0
D^{*0}$ threshold and its width is extremely narrow ($<1.2$ MeV).
Later LHCb Collaboration determined its $J^{PC}=1^{++}$
\cite{R.Aaij:2013}. Many theoretical groups interpreted $X(3872)$ as
the molecular candidate of the $\bar{D}D^*$ system
\cite{C.Y.Wong:2004,E.S.Swanson:2004,M.Suzuki:2005,S.L.Zhu:2008}.

Besides $X(3872)$, a family of so called $Y$ states were also
reported. $Y(4260)$ was observed by $BARBAR$ Collaboration in the
invariant mass spectrum of $\pi^+ \pi^- J/\psi$ in the initial-state
radiation process $e^+ e^- \rightarrow \gamma_{ISR} \pi^+ \pi^-
J/\psi$ \cite{B. Aubert:2005}. Later Belle Collaboration observed a
peak near $4.25$ GeV and a new structure around $4.05$ GeV which was
denoted later as $Y(4008)$ \cite{C.Z. Yuan:2007, Z.Q. Liu:2013}.
$Y(4360)$ was observed in the reaction $e^+ e^- \rightarrow \pi^+
\pi^- \psi(2S)$ by $BARBAR$ \cite{B. Aubert:2007}. Almost at the
same time, Belle observed two resonant structures in the $\pi^+
\pi^- \psi(2S)$ invariant mass distribution $Y(4360)$ and $Y(4660)$
\cite{X.L. Wang:2007}, which was confirmed by $BARBAR$ via the
initial-state radiation process $e^+ e^- \rightarrow \pi^+ \pi^-
\psi(2S)$ \cite{J.P. Lees:2014}. $Y(4630)$ was reported as a
near-threshold enhancement in the $e^+ e^- \rightarrow \Lambda_{c}^+
\Lambda_{c}^-$ process \cite{G.Pakhlova:2008}.

The group of charged charmonium-like and bottomonium-like states are
even more exotic. The lightest charged charmonium-like state
$Z_c(3900)$ was observed in the $J/\psi \pi^{\pm}$ invariant mass in
the process $Y(4260) \rightarrow J/\psi \pi^+ \pi^-$ by BESIII
Collaboration \cite{Ablikim:2013}, by Belle Collaboration with ISR
\cite{Z.Q. Liu:2013-2} and by using CLEO data \cite{T.Xiao:2013}.
Its decay mode implies that $Z_c(3900)$ is a hidden-charm structure.
$Z_c(4025)$ was observed in the $\pi^{\mp}$ recoil mass spectrum in
the process $e^+ e^- \rightarrow (D^* \bar{D}^*)^{\pm} \pi^{\mp}$
\cite{Ablikim:2014}. $Z_c(4020)$ was reported in the $\pi^{\pm} h_c$
mass spectrum in the process $e^+ e^- \rightarrow \pi^+ \pi^- h_c$
\cite{M. Ablikim:2013}. Moreover, two charged bottomonium-like
resonances $Z_b(10610)$ and $Z_b(10650)$ were observed in the
$\pi^{\pm} \Upsilon(nS)$ and $\pi^{\pm} h_b$ mass spectrum in the
$\Upsilon(5S)$ decays \cite{I.Adachi:2011}. $Z_1(4050)$ and
$Z_2(4250)$ were observed in the $\pi^+ \chi_{c1}$ invariant mass
distribution in the $\bar{B}^0 \rightarrow K^- \pi^+ \chi_{c1}$
decays \cite{R.Mizuk:2008}. $Z_c(4485)$ was observed by Belle
Collaboration in the $\pi^{\pm} \psi^{\prime}$ invariant mass
distribution in the exclusive $B \rightarrow K \pi^{\pm}
\psi^{\prime}$ decays \cite{Choi:2008}. Later its spin and parity
were determined as $J^P = 1^+$ \cite{K.Chilikin:2013}. The
charmonium-like state $Z_c(4200)$ was observed in the $J/\psi \pi^+$
mode with a significance of $8.2\sigma$ when performing the
amplitude analysis of $B \rightarrow J/\psi K \pi$ \cite{zc4200}.

These $XYZ$ states either decay into one charmonium/bottomonium
state plus light mesons or into a pair of open-charm/open-bottom
heavy mesons. Many of them do not fit into the conventional $q\bar
q$ meson spectrum in the quark model. Some of them were interpreted
as the candidates of the hybrid meson \cite{ZHUPLB2005}, molecular
states \cite{F.E.Close:2004, M.B.Voloshin:2004, C.Y.Wong:2004,
E.S.Swanson:2004, N.A.Tornqvist:2004, Y.R.Liu:2010,lining},
tetraquark states \cite{H.Hogaasen:2006, D.Ebert:2006,
N.Barnea:2006, Y.Cui:2007, R.D.Matheus:2007, T.W.Chiu:2007} and so
on. For example, $Z_c(3900)$ was interpreted as the isovector axial
vector molecular partner of $X(3872)$
\cite{Q.Wang:2013,Oset,zhaolu}. Similarly $Z_c(4025)$ was speculated
to be the $D^\ast D^\ast$ molecular candidate
\cite{chenwei1,hejun,sun}. There are also some other speculations
about their nature \cite{C.D. Deng:2014, J.M. Dias:2014}.
$Z_b(10610)$ and $Z_b(10650)$ are generally regarded as the
candidates of the $\bar{B}B^*$ and $\bar{B}^*B^*$ molecular states
\cite{Y.R.Liu:2008, X.Liu:2009, A.E.Bondar:2011,liuxiang1}.

However, it is not very natural to explain  $Z_c(4200)$ and
$Z_c(4485)$ as the S-wave molecular states composed of two S-wave
heavy mesons. Instead, $Z_c(4485)$ was proposed as the cousin
molecular state of $Z_c(3900)$ and $Z_c(4025)$ composed of
$D(D^\ast)$ and its radial excitation \cite{mali,liuxiaohai}.

Another interesting possibility is that some charged $Z_c$ states
might be tetraquark candidates. The light $q\bar{q}q\bar{q}$
tetraquark system was first studied in the MIT bag model \cite{R.L.
Jaffe:1977-1, R.L. Jaffe:1977-2}, where the multiquark mass spectrum
mostly depend on the chromomagnetic interaction among the quarks.
When considering the chromomagnetic interaction, it is convenient to
adopt the $SU(6)_{cs}$ representation which is the eigenstate of the
color-magnetic (CM) interaction and can be constructed as the direct
product of the $SU(3)$ color and the $SU(2)$ spin group. The bag
model was later used to discuss the hidden-charm/bottom tetraquark
system \cite{K.T. Chao:1980, K.T. Chao:1981}. The hidden-charm
tetraquarks were also studied in the constituent quark model (CQM)
\cite{Y.Cui:2007, Y. Cui:2006}.

In this work we will investigate whether some of the charged $Z_c$
states could be the tetraquark candidates. We will discuss the
mixing of the hidden-charm tetraquark states in the different
color-spin representation and possible mass splitting of the
hidden-charm tetraquark states in the framework of the
chromomagnetic interaction. We will employ two schemes to fix the
strength of the CM interaction and extract the masses and wave
functions of the $J^P=1^+, 0^+, 2^+$ tetraquark systems. Then we
compare the hidden-charm tetraquark spectrum with the current
experimental data.

The paper is organized as follows. After the introduction, we
present the chromomagnetic hamiltonian and the tetraquark model in
Section \ref{potential}. In Section \ref{mass}, we discuss the
masses of the possible tetraquark candidates. We explore the decay
pattern of the tetraquark system in Section \ref{decay}. The last
section is the discussion and summary.

\section{Heavy Tetraquark}\label{potential}

\subsection{The Chromomagnetic Hamiltonian}

For the tetraquark system, we consider the chromomagnetic (CM)
interaction to derive the mass splitting. The Hamiltonian reads
\begin{equation}
H = \sum_{i}m_{i} + H_{CM} \label{Hamiltonian}
\end{equation}
where $m_{i}$ is the mass of the $i$-th constituent quark. $H_{CM}$
describes the CM interaction which is derived from one gluon
exchange \cite{R.L. Jaffe:1977-1, R.L. Jaffe:1977-2, R.L.
Jaffe:1977-3, T. Degrand:1975}
\begin{equation}
H_{CM} = -\sum_{i>j}v_{ij}\vec{\lambda}_{i}\cdot\vec{\lambda}_{j}
\vec{\sigma}_{i}\cdot\vec{\sigma}_{j} \label{CM-Hamiltonian}
\end{equation}
where $\vec{\lambda}_{i}$ is the quark color operator and
$\vec{\sigma}_{i}$ is the spin operator. For the anti-quark,
$\vec{\lambda}_{\bar{q}} = -\vec{\lambda}_{q}^{*}$ and
$\vec{\sigma}_{\bar{q}} = -\vec{\sigma}_{q}^{*}$. $v_{ij}$
represents the interaction strength between two quarks. Therefore,
$v_{ij}$ depends on the wavefunction of the multiquark system. For
example, $v_{ij}$ takes different values in the $q\bar{q}$, $qqq$
and $q\bar{q}q\bar{q}$ systems. In the bag model, $v_{ij}$ depends
on the bag radius and the constituent quark mass. On the other hand,
the constituent quark model (CQM) is very successful in describing
the meson and baryon spectrum, where the color-magnetic interaction
leads to the mass splitting between the octet and decuplet baryons.
We follow the CQM convention and adopt $v_{ij} = v \frac{m^2_u}{m_i
m_j}$. The parameter $v$ depends on the multiquark system.

\subsection{Hidden-charm tetraquark wavefunction}

For the $qq\bar{q}\bar{q}$ tetraquark system, the CM wavefunction
can be constructed either as $qq \otimes \bar{q}\bar{q}$ or
$q\bar{q} \otimes q\bar{q}$. We use $Q$, $\bar{Q}$ and $\tilde{Q}$
to represent the configuration $qq$, $\bar{q}\bar{q}$ and $q\bar{q}$
respectively. We use the notation $|D_6, D_{3c}, S, N \rangle$ to
represent the diquark configuration, where $D_6$, $D_{3c}$, $S$ and
$N$ are the $SU(6)$ color-spin coupling representations, $SU(3)_c$
color representations, spin and number of the constituent quarks
respectively. Based on the $SU(6)_{cs} \supset SU(3)_{c} \otimes
SU(2)_{s}$ group theory, there are four types of representations for
the diquark $qq$: $| 21, \bar{3}_c, 0, 2 \rangle$, $| 21, 6_c, 1, 2
\rangle$, $| 15, \bar{3}_c, 1, 2 \rangle$ and $| 15, \bar{6}_c, 0, 2
\rangle$. For the anti-diquark $\bar{q}\bar{q}$, there are also four
types of representations : $| \bar{21}, 3_c, 0, 2 \rangle$, $|
\bar{21}, \bar{6}_c, 1, 2 \rangle$, $| \bar{15}, 3_c, 1, 2 \rangle$
and $| \bar{15}, 6_c, 0, 2 \rangle$. For the $q\bar{q}$ system,
there are also four types of representations : $| 1, 1_c, 0, 2
\rangle$, $| 35, 1_c, 1, 2 \rangle$, $| 35, 8_c, 1, 2 \rangle$ and
$| 35, 8_c, 0, 2 \rangle$.

For the tetraquark system $q_1 q_2 \bar{q}_3 \bar{q}_4$ with four
different flavors, the CM interaction matrix element between two
$SU(6)_{cs}$ eigenstates $| k \rangle$ and $| l \rangle$ is
\begin{eqnarray}
V_{CM}(q_{1}q_{2}\bar{q}_{3}\bar{q_{4}}) &=& \langle k |H_{CM}| l
\rangle = V_{12}(q_{1}q_{2}) + V_{13}(q_{1}\bar{q_{3}}) \nonumber\\
& + & V_{14}(q_{1}\bar{q_{4}}) + V_{23}(q_{2}\bar{q_{3}}) +
V_{24}(q_{2}\bar{q_{4}}) \nonumber\\ & + &
V_{34}(\bar{q_{3}}\bar{q_{4}}) \label{CM-interaction}
\end{eqnarray}
where
\begin{equation}
V_{ij}(Q) =
v_{ij}\vec{\lambda}_{i}\cdot\vec{\lambda}_{j}
\vec{\sigma}_{i}\cdot\vec{\sigma}_{i} =
-\frac{v_{ij}}{2}[\bar{C}(Q)-16N] \label{diquark-interaction}
\end{equation}
and
\begin{equation}
V_{ij}(\tilde{Q}) =
v_{ij}\vec{\lambda}_{i}\cdot\vec{\lambda}^{*}_{j}
\vec{\sigma}_{i}\cdot\vec{\sigma}^{*}_{j} =
\frac{v_{ij}}{2}[\bar{C}(\tilde{Q})-16N]
\end{equation}
For the diquark system we have
\begin{equation}
\bar{C}(Q) = \bar{C}(\tilde{Q}) = C_6 - C_3 - \frac{8}{3}S(S + 1)
\end{equation}
where $C_6$ and $C_3$ are the Casimir operators of $SU(6)_{cs}$ and
$SU(3)_c$ groups. $S$ is the spin operator.

Based on the $SU(6)_{cs}$ group decomposition, the color-spin
wavefunction of the $J^p=1^+$ tetraquark $SU(6)_{cs}$ eigenstates
can be constructed in the $Q \otimes \bar{Q}$ form
\begin{eqnarray}
| 35, 1_c, 1, 4 \rangle = | 21, 6_c, 1, 2 \rangle \otimes |
\bar{21}, \bar{6}_c, 1, 2 \rangle \label{qq-base1}
\end{eqnarray}
\begin{eqnarray}
| 35, 1_c, 1, 4 \rangle = | 15, \bar{3}_c, 1, 2 \rangle \otimes |
\bar{15}, 3_c, 1, 2 \rangle \label{qq-base2}
\end{eqnarray}
\begin{eqnarray}
| 35, 1_c, 1, 4 \rangle &=& \sqrt{\frac{1}{3}} | 21, \bar{3}_c, 0, 2
\rangle \otimes | \bar{15}, 3_c, 1, 2 \rangle \nonumber\\ &-&
\sqrt{\frac{2}{3}} | 21, 6_c, 1, 2 \rangle \otimes | \bar{15},
\bar{6}_c, 1, 2 \rangle \label{qq-base3}
\end{eqnarray}
\begin{eqnarray}
| 280, 1_c, 1, 4 \rangle &=& \sqrt{\frac{2}{3}} | 21, \bar{3}_c, 0,
2 \rangle \otimes | \bar{15}, 3_c, 1, 2 \rangle \nonumber\\ &+&
\sqrt{\frac{1}{3}} | 21, 6_c, 1, 2 \rangle \otimes | \bar{15},
\bar{6}_c, 1, 2 \rangle \label{qq-base4}
\end{eqnarray}
\begin{eqnarray}
| 35, 1_c, 1, 4 \rangle &=& \sqrt{\frac{1}{3}} | \bar{21}, 3_c, 0, 2
\rangle \otimes | 15, \bar{3}_c, 1, 2 \rangle \nonumber\\ &-&
\sqrt{\frac{2}{3}} | \bar{21}, \bar{6}_c, 1, 2 \rangle \otimes | 15,
6_c, 1, 2 \rangle \label{qq-base5}
\end{eqnarray}
\begin{eqnarray}
| 280, 1_c, 1, 4 \rangle &=& \sqrt{\frac{2}{3}} | \bar{21}, 3_c, 0,
2 \rangle \otimes | 15, \bar{3}_c, 1, 2 \rangle \nonumber\\ &+&
\sqrt{\frac{1}{3}} | \bar{21}, \bar{6}_c, 1, 2 \rangle \otimes | 15,
6_c, 1, 2 \rangle \label{qq-base6}
\end{eqnarray}

The CM wavefunctions of the $J^P=0^+$ and $J^P=2^+$ tetraquark
states are listed in the appendix. These wavefunctions are the
eigenstates of the CM interaction $V_{ij}(Q)$ and $V_{ij}(\bar{Q})$.
The CM interaction $V_{CM}$ also has the form $V_{ij}(\tilde{Q})$.
In order to get their eigenstates, we need to do the recoupling from
$Q \otimes \bar{Q}$ to $\tilde{Q} \otimes \tilde{Q}$. Based on
Wigner and Racah coefficients of $SU(6)_{cs} \supset SU(3)_{c}
\otimes SU(2)_{s}$ \cite{D. Strottman:1976, S. I. So:1976}, the
$1^+$ $SU(6)_{cs}$ eigenstates in terms of $q_1 \bar{q}_3\otimes q_2
\bar{q}_4$ are
\begin{eqnarray}
| 35, 1_c, 1, 4 \rangle &=& \frac{\sqrt{3}}{3} | q_1 \bar{q}_3 1,
1_c, 0, 2 \rangle \otimes | q_2 \bar{q}_4 35, 1_c, 1, 2 \rangle
\nonumber\\ &+& \frac{\sqrt{6}}{6} | q_1 \bar{q}_3 35, 8_c, 0, 2
\rangle \otimes | q_2 \bar{q}_4 35, 8_c, 1, 2 \rangle \nonumber\\
&+& \frac{\sqrt{3}}{3} | q_1 \bar{q}_3 35, 1_c, 1, 2 \rangle \otimes
| q_2 \bar{q}_4 1, 1_c, 0, 2 \rangle \nonumber\\ &+&
\frac{\sqrt{6}}{6} | q_1 \bar{q}_3 35, 8_c, 1, 2 \rangle \otimes |
q_2 \bar{q}_4 35, 8_c, 0, 2 \rangle \label{qqbar-base1}
\end{eqnarray}
\begin{eqnarray}
| 35, 1_c, 1, 4 \rangle &=& \frac{\sqrt{6}}{6} | q_1 \bar{q}_3 1,
1_c, 0, 2 \rangle \otimes | q_2 \bar{q}_4 35, 1_c, 1, 2 \rangle
\nonumber\\ &-& \frac{\sqrt{3}}{3} | q_1 \bar{q}_3 35, 8_c, 0, 2
\rangle \otimes | q_2 \bar{q}_4 35, 8_c, 1, 2 \rangle \nonumber\\
&+& \frac{\sqrt{6}}{6} | q_1 \bar{q}_3 35, 1_c, 1, 2 \rangle \otimes
| q_2 \bar{q}_4 1, 1_c, 0, 2 \rangle \nonumber\\ &-&
\frac{\sqrt{3}}{3} | q_1 \bar{q}_3 35, 8_c, 1, 2 \rangle \otimes |
q_2 \bar{q}_4 35, 8_c, 0, 2 \rangle \label{qqbar-base2}
\end{eqnarray}
\begin{eqnarray}
| 35, 1_c, 1, 4 \rangle &=& \frac{1}{2} | q_1 \bar{q}_3 1, 1_c, 0, 2
\rangle \otimes | q_2 \bar{q}_4 35, 1_c, 1, 2 \rangle \nonumber\\
&-& \frac{1}{2} | q_1 \bar{q}_3 35, 1_c, 1, 2 \rangle \otimes | q_2
\bar{q}_4 1, 1_c, 0, 2 \rangle \nonumber\\ &-& \frac{2}{3} | q_1
\bar{q}_3 35, 8_c, 1, 2 \rangle \otimes | q_2 \bar{q}_4 35, 8_c, 1,
2 \rangle \nonumber\\ &-& \frac{\sqrt{2}}{6} | q_1 \bar{q}_3 35,
1_c, 1, 2 \rangle \otimes | q_2 \bar{q}_4 35, 1_c, 1, 2 \rangle
\label{qqbar-base3}
\end{eqnarray}
\begin{eqnarray}
| 280, 1_c, 1, 4 \rangle &=& -\frac{1}{2} | q_1 \bar{q}_3 35, 8_c,
0, 2 \rangle \otimes | q_2 \bar{q}_4 35, 8_c, 1, 2 \rangle
\nonumber\\ &+& \frac{1}{2} | q_1 \bar{q}_3 35, 8_c, 1, 2 \rangle
\otimes | q_2 \bar{q}_4 35, 8_c, 0, 2 \rangle \nonumber\\ &-&
\frac{\sqrt{2}}{6} | q_1 \bar{q}_3 35, 8_c, 1, 2 \rangle \otimes |
q_2 \bar{q}_4 35, 8_c, 1, 2 \rangle \nonumber\\ &+& \frac{2}{3} |
q_1 \bar{q}_3 35, 1_c, 1, 2 \rangle \otimes | q_2 \bar{q}_4 35, 1_c,
1, 2 \rangle \label{qqbar-base4}
\end{eqnarray}
\begin{eqnarray}
| 35, 1_c, 1, 4 \rangle &=& -\frac{1}{2} | q_1 \bar{q}_3 1, 1_c, 0,
2 \rangle \otimes | q_2 \bar{q}_4 35, 1_c, 1, 2 \rangle \nonumber\\
&+& \frac{1}{2} | q_1 \bar{q}_3 35, 1_c, 1, 2 \rangle \otimes | q_2
\bar{q}_4 1, 1_c, 0, 2 \rangle \nonumber\\ &-& \frac{2}{3} | q_1
\bar{q}_3 35, 8_c, 1, 2 \rangle \otimes | q_2 \bar{q}_4 35, 8_c, 1,
2 \rangle \nonumber\\ &-& \frac{\sqrt{2}}{6} | q_1 \bar{q}_3 35,
1_c, 1, 2 \rangle \otimes | q_2 \bar{q}_4 35, 1_c, 1, 2 \rangle
\label{qqbar-base5}
\end{eqnarray}
\begin{eqnarray}
| 280, 1_c, 1, 4 \rangle &=& \frac{1}{2} | q_1 \bar{q}_3 35, 8_c, 0,
2 \rangle \otimes | q_2 \bar{q}_4 35, 8_c, 1, 2 \rangle \nonumber\\
&-& \frac{1}{2} | q_1 \bar{q}_3 35, 8_c, 1, 2 \rangle \otimes | q_2
\bar{q}_4 35, 8_c, 0, 2 \rangle \nonumber\\ &-& \frac{\sqrt{2}}{6} |
q_1 \bar{q}_3 35, 8_c, 1, 2 \rangle \otimes | q_2 \bar{q}_4 35, 8_c,
1, 2 \rangle \nonumber\\ &+& \frac{2}{3} | q_1 \bar{q}_3 35, 1_c, 1,
2 \rangle \otimes | q_2 \bar{q}_4 35, 1_c, 1, 2 \rangle
\label{qqbar-base6}
\end{eqnarray}
According to the $SU(3)_c$ and $SU(2)_s$ symmetry, the $SU(6)_{cs}$
eigenstates in terms of $q_2 \bar{q}_3 \otimes q_1 \bar{q}_4$ have
the same form with those of $q_1 \bar{q}_3 \otimes q_2 \bar{q}_4$ in
Eqs. (\ref{qqbar-base1}), (\ref{qqbar-base3}) and
(\ref{qqbar-base4}). There appears an extra minus sign in the
tetraquark states in Eqs. (\ref{qqbar-base2}), (\ref{qqbar-base5})
and (\ref{qqbar-base6}) when we change the basis from $q_2 \bar{q}_3
\otimes q_1 \bar{q}_4$ to $q_1 \bar{q}_3 \otimes q_2 \bar{q}_4$. For
the $J^p=0^+$ and $J^P=2^+$ tetraquark states, the $SU(6)_cs$
eigenstates in the form of $\tilde{Q} \otimes \tilde{Q}$ are listed
in the appendix.

Using the above $SU(6)_{cs}$ eigenstates in Eqs.
(\ref{qq-base1})-(\ref{qqbar-base6}), we can calculate each
individual term in Eq. (\ref{CM-interaction}), obtain the
eigenvalues of the CM interaction matrix $V_{CM}$, and derive the
wave function and mass of the tetraquark system.

\section{Possible tetraquark candidates among various $Z_c$ states}
\label{mass}

In order to extract the tetraquark mass, we need the values of the
constituent quark mass and the parameter $v$. Recall that the
charmonium $J/\psi$ and $\eta_c$ can be treated as the $SU(6)_{cs}$
diquark $c\bar{c}$ state $|35, 1_c, 1, 2\rangle$ and $|1, 1_c, 0,
2\rangle$. Similarly, the charmed mesons $D^*$ and $D$ can be
treated as $SU(6)_{cs}$ diquark $c\bar{u}$ state $|35, 1_c, 1,
2\rangle$ and $|1, 1_c, 0, 2\rangle$. With Eq.
(\ref{diquark-interaction}) and the meson masses from PDG
\cite{PDG}, we can extract the masses of the $u$, $c$, $s$ and $b$
constituent quarks.
\begin{eqnarray}
\left\{
  \begin{array}{ll}
     & \hbox{$M(J/\psi) = 2m_c + \frac{16}{3} v_{c\bar{c}} (\frac{m_u}{m_c})^2 $;} \\
     & \hbox{$M(\eta_c) = 2m_c - 16v_{c\bar{c}} (\frac{m_u}{m_c})^2 $;} \\
     & \hbox{$M(D^*) = m_u + m_c + \frac{16}{3} v_{c\bar{u}} \frac{m_u}{m_c} $;} \\
     & \hbox{$M(D) =  m_u + m_c - 16v_{c\bar{u}} \frac{m_u}{m_c} $;} \\
     & \hbox{$M(D^*_s) =  m_s + m_c + \frac{16}{3} v_{c\bar{s}} \frac{m_u}{m_c} \frac{m_u}{m_s} $;} \\
     & \hbox{$M(D_s) =  m_s + m_c - 16v_{c\bar{s}} \frac{m_u}{m_c} \frac{m_u}{m_s} $;} \\
     & \hbox{$M(\Upsilon) =  2m_b + 16v_{b\bar{b}} (\frac{m_u}{m_b})^2 \approx 2m_b $.} \\
  \end{array}
\right.
\end{eqnarray}
From the above equation, we get
\begin{eqnarray}
\left\{
  \begin{array}{ll}
     & \hbox{$m_c = 1534 $ MeV;} \\
     & \hbox{$m_u = 437 $ MeV;} \\
     & \hbox{$m_s = 542 $ MeV;} \\
     & \hbox{$m_b = 4730 $ MeV;} \\
  \end{array}
\right.
\end{eqnarray}
According Eq. (\ref{CM-interaction}), $V_{CM}(qc\bar{q}\bar{c})$
reads
\begin{eqnarray}
V_{CM}(qc\bar{q}\bar{c}) &=& \frac{m_u}{m_c} V_{12} + V_{13} +
\frac{m_u}{m_c} V_{14} \nonumber\\ &+& \frac{m_u}{m_c} V_{23} +
(\frac{m_u}{m_c})^2 V_{24} + \frac{m_u}{m_c} V_{34}
\label{qcqc-interaction}
\end{eqnarray}
After diagonalizing the mass matrix $V_{CM}$ for the $J^P=1^+$
$qc\bar{q}\bar{c}$ tetraquark states, we get six eigenvalues:
$-15.9v$, $-4.1v$, $-1.5v$, $1.7v$, $5.6v$, $5.8v$ which are listed
in Table \ref{tab:Vcm-eigenvalue}. Sometimes we use the eigenvalues
to denote the state. In the following we discuss two schemes to fix
the parameter $v$ and extract the tetraquark spectrum.

\subsection{Scheme I: Using the mass of one of the $Z_c$ states as
input}\label{scheme1}

Assuming that $Z_c(3900)$ is one of the six tetraquark states, the
parameter $v$ can be fixed. Similarly, $Z_c(4025)$ and $Z_c(4200)$
can also be used as input to extract the value of $v$. Throughout
our discussion, we require $v$ to be positive. Then we use the
obtained $v$ to calculate the masses of the other eigenstates, which
are listed in Table \ref{tab:v-mass}.

If $Z_c(3900)$ is appointed as the state with the eigenvalue
$-15.9v$, $-4.1v$ and $-1.5v$, it is quite difficult to accommodate
either $Z_c(4025)$ or $Z_c(4200)$ among the six states. If
$Z_c(4025)$ is appointed as the state with the eigenvalue $1.7v$,
the mass of the state with the eigenvalue $5.6v$ is $4215.4$ MeV
which is close to $Z_c(4200)$, while the mass of the state with the
eigenvalue $-1.5v$ is $3868.8$ MeV which is 30 MeV lower than
$Z_c(3900)$. Unfortunately, the lowest axial vector tetraquark state
is around 3166 MeV. Such a scheme is not realistic.

If $Z_c(4200)$ is appointed as the state with the eigenvalue $5.6v$,
the mass of the state with the eigenvalue $1.7v$ is $4020.3$ MeV
which is close to $Z_c(4020)$. The mass of the state with the
eigenvalue $-1.5v$ is $3872.9$ MeV, which is 28 MeV lower than
$Z_c(3900)$. In this case the lowest state is around 3210 MeV, which
is also quite unrealistic. It's almost impossible to accommodate all
the three charged states $Z_c(3900)$, $Z_c(4025)$ and $Z_c(4200)$
within the axial vector tetraquark spectrum simultaneously. In other
words, at least one or two of these states is not a tetraquark
candidate.

\begin{table}[htbp]
\caption{The masses of the six axial vector $qc\bar{q}\bar{c}$
tetraquark states when the parameter $v$ is fixed by the mass of
$Z_c(3900)$, $Z_c(4025)$ and $Z_c(4200)$. The eigenvalue is used to
denote the state as the subscript.} \label{tab:v-mass}
\begin{center}
\begin{tabular}{c  c  c  c  c  c  c  c  }
\hline \hline  $v$ & $-15.9v$ & $-4.1v$ & $-1.5v$ & $1.7v$ & $5.6v$ & $5.8v$ \\
\hline
$v_{-15.9}^{Z_c(3900)} = 2.6$ & $3900$   &  $3931.2$  &  $3938.0$  &  $3946.5$ &  $3956.8$  &  $3957.3$\\
\hline
$v_{-4.1}^{Z_c(3900)} = 10.2$ & $3779.1$   &  $3900$  &  $3926.6$  &  $3959.4$ &  $3999.4$  &  $4001.4$\\
\hline
$v_{-1.5}^{Z_c(3900)} = 28.0$ & $3496.8$  &  $3827.2$  &  $3900$  &  $3989.6$  &  $4098.8$  &  $4104.4$\\
\hline
$v_{1.7}^{Z_c(4025)}= 48.8$ & $3165.7$  &  $3741.8$  &  $3868.8$  &  $4025$  &  $4215.4$  &  $4225.2$\\
\hline
$v_{5.6}^{Z_c(4025)} = 14.8$ & $3706.3$  &  $3881.2$  &  $3919.8$  &  $3967.2$  &  $4025$  &  $4028.0$\\
\hline
$v_{5.8}^{Z_c(4025)} = 14.3$ & $3714.5$  &  $3883.3$  &  $3920.5$  &  $3966.3$  &  $4022.1$  &  $4025$\\
\hline
$v_{5.6}^{Z_c(4200)} = 46.1$ & $3209.4$  &  $3753.1$  &  $3872.9$  &  $4020.3$  &  $4200$  &  $4209.2$\\
\hline
$v_{5.8}^{Z_c(4200)} = 44.5$ & $3234.7$  &  $3759.6$  &  $3875.3$  &  $4017.6$  &  $4191.1$  &  $4200$\\
\hline\hline
\end{tabular}
\end{center}
\end{table}

\subsection{Scheme II: using the mass splitting of two $Z_c$ states as
input}\label{scheme2}

The parameter $v$ can be extracted from the mass splitting if we
assume two of the three states $Z_c(3900)$, $Z_c(4025)$ and
$Z_c(4200)$ are the $1^+$ $qc\bar{q}\bar{c}$ tetraquark states. As
pointed out in Section \ref{decay}, the state with the eigenvalue
$-4.1v$ does not decay to $J/\psi \pi$. Thus it is not appropriate
to assign it as $Z_c(3900)$. Therefore we only assume $Z_c(3900)$ as
the state either with the eigenvalue $-15.9$ or $-1.5v$. Once the
value of $v$ is extracted, we obtain the whole spectrum. The results
are listed in Table \ref{tab:v-splitting-mass1}.

If $Z_c(3900)$ and $Z_c(4025)$ are assigned as the state with the
eigenvalue $-1.5v$ and $1.7v$ respectively, the resulting mass of
the state with the eigenvalue $5.6v$ is $4177.3$ MeV, which is close
to $Z_c(4200)$. Unfortunately the lowest state is around 3338 MeV,
which is unrealistic. Similarly, if $Z_c(3900)$ and $Z_c(4200)$ are
assigned as the state with the eigenvalue $-1.5v$ and $5.6v$
respectively, the mass of the state with the eigenvalue $1.7v$ is
$4035.2$ MeV, which is close to $Z_c(4025)$. If $Z_c(4025)$ and
$Z_c(4200)$ are treated as the state with the eigenvalue $1.7v$ and
$5.6v$ respectively, the mass of the state with the eigenvalue
$-1.5v$ is $3881.4$ MeV, which is close to $Z_c(3900)$. Now the
lowest state is around 3235 MeV. Although we could accommodate all
three charged states $Z_c(3900)$, $Z_c(4025)$ and $Z_c(4200)$ as the
axial vector tetraquark candidates, the resulting mass of the lowest
state is always too low and unrealistic. In other words, not all
these three states are tetraquark candidates, which is consistent
with the conclusion in the previous subsection.

\begin{table}[htbp]
\caption{The masses of the $1^+$ $qc\bar{q}\bar{c}$ tetraquark
states when the parameter $v$ is fixed by the mass difference of two
$Z_c$ states.} \label{tab:v-splitting-mass1}
\begin{center}
\begin{tabular}{ c  c  c  c  c  c    }
\hline \hline  $-15.9v$ & $-4.1v$ & $-1.5v$ & $1.7v$ & $5.6v$ & $5.8v$\\
\hline \hline
\multicolumn{6}{c} {$Z_c(3900) \rightarrow -15.9v, Z_c(4025) \rightarrow -4.1v, v = 10.6 $} \\
\hline
$3900$  &  $4025$  &  $4052.5$  &  $4086.4$  &  $4127.8$  &  $4129.9$ \\
\hline\hline
\multicolumn{6}{c} {$Z_c(3900) \rightarrow -15.9v, Z_c(4025) \rightarrow -1.5v, v = 8.7 $} \\
\hline
$3900$  &  $4002.4$  &  $4025$  &  $4052.8$  &  $4086.6$  &  $4088.4$\\
\hline\hline
\multicolumn{6}{c} {$Z_c(3900) \rightarrow -15.9v, Z_c(4025) \rightarrow 1.7v, v = 7.1 $} \\
\hline
$3900$  &  $3983.8$  &  $4002.3$  &  $4025$  &  $4052.7$  &  $4054.1$\\
\hline\hline
\multicolumn{6}{c} {$Z_c(3900) \rightarrow -15.9v, Z_c(4025) \rightarrow 5.6v, v = 5.81 $} \\
\hline
$3900$  &  $3968.6$  &  $3983.7$  &  $4002.3$  &  $4025$  &  $4026.2$\\
\hline\hline
\multicolumn{6}{c} {$Z_c(3900) \rightarrow -15.9v, Z_c(4025) \rightarrow 5.8v, v = 5.76 $} \\
\hline
$3900$  &  $3968$  &  $3983$  &  $4001.4$  &  $4023.9$  &  $4025$\\
\hline\hline
\multicolumn{6}{c} {$Z_c(3900) \rightarrow -1.5v, Z_c(4025) \rightarrow 1.7v, v = 39.1 $} \\
\hline
$3337.5$  &  $3798.4$  &  $3900$  &  $4025$  &  $4177.3$  &  $4185.2$\\
\hline\hline
\multicolumn{6}{c} {$Z_c(3900) \rightarrow -1.5v, Z_c(4025) \rightarrow 5.6v, v = 17.6 $} \\
\hline
$3646.5$  &  $3854.2$  &  $3900$  &  $3956.3$  &  $4025$  &  $4028.5$\\
\hline\hline
\multicolumn{6}{c} {$Z_c(4025) \rightarrow 1.7v, Z_c(4200) \rightarrow 5.6v, v = 44.9 $} \\
\hline
$3235.3$  &  $3764.7$  &  $3881.4$  &  $4025$  &  $4200$  &  $4209$\\
\hline\hline
\multicolumn{6}{c} {$Z_c(4025) \rightarrow 1.7v, Z_c(4200) \rightarrow 5.8v, v = 42.7 $} \\
\hline
$3273.8$  &  $3777.4$  &  $3888.4$  &  $4025$  &  $4191.5$  &  $4200$\\
\hline\hline
\multicolumn{6}{c} {$Z_c(3900) \rightarrow -1.5v, Z_c(4200) \rightarrow 5.6v, v = 42.3 $} \\
\hline
$3291.6$  &  $3790.1$  &  $3900$  &  $4035.2$  &  $4200$  &  $4208.5$\\
\hline\hline
\end{tabular}
\end{center}
\end{table}

\subsection{The $qc\bar{s}\bar{c}$, $sc\bar{s}\bar{c}$ and hidden-bottom tetraquark states}

We assume $Z_c(4025)$ as the $qc\bar{q}\bar{c}$ tetraquark state
with the eigenvalue $1.7v$ to fix the parameter $v$ and collect the
numerical results for the $qc\bar{s}\bar{c}$ and $sc\bar{s}\bar{c}$
tetraquark states in Table \ref{tab:mass-s}.

\begin{table}[htbp]
\caption{The masses of the $qc\bar{q}\bar{c}$, $qc\bar{s}\bar{c}$
and $sc\bar{s}\bar{c}$ tetraquark states with $J^P=0^+, 1^+, 2^+$.
The parameter $v$ is fixed assuming $Z(4025)$ as the tetraquark
state with the eigenvalue $1.7v$ } \label{tab:mass-s}
\begin{center}
\begin{tabular}{c| c  | c c c }
\hline \hline   &     &  $0^+$ & $1^+$ & $2^+$   \\
\hline
\multirow{4}{*}{$qc\bar{q}\bar{c}$} & $V_{CM}$ &   $-18.6v$ & $-15.9v$ & $2.7v$   \\
\cline{2-5}
                                    & $M$(MeV) &   $3033.9$ & $3165.7$ & $4073.8$   \\
\cline{2-5}
                                    & $V_{CM}$ &   $-7.4v$ & $-4.1v$   & $5.8v$   \\
\cline{2-5}
                                    & $M$(MeV) &   $3580.7$ & $3741.8$ & $4225.2$   \\
\hline\hline
\multirow{4}{*}{$qc\bar{s}\bar{c}$} & $V_{CM}$ &   $-15.6v$ & $-12.8v$ & $2.5v$   \\
\cline{2-5}
                                    & $M$(MeV) &   $3285.4$ & $3422.1$ & $4169.1$   \\
\cline{2-5}
                                    & $V_{CM}$ &   $-6.5v$  & $-3.8v$   & $4.7v$   \\
\cline{2-5}
                                    & $M$(MeV) &   $3729.7$ & $3861.5$ & $4276.5$   \\
\hline\hline
\multirow{2}{*}{$sc\bar{s}\bar{c}$} & $V_{CM}$ &   $-13.1v$ & $-10.3v$ & $2.3v$   \\
\cline{2-5}
                                    & $M$(MeV) &   $3512.4$ & $3649.1$ & $4264.3$   \\
\cline{2-5}
                                    & $V_{CM}$ &   $-5.7v$  & $-3.2v$   & $3.9v$   \\
\cline{2-5}
                                    & $M$(MeV) &   $3873.7$ & $3995.8$ & $4342.4$   \\
\hline \hline
\end{tabular}
\end{center}
\end{table}

We extend the same formalism to investigate the hidden-bottom
tetraquark states. The results are collected in Tables
\ref{tab:Vcm-eigenvalue-b}, \ref{tab:v-mass-b}, \ref{tab:mass-s-b}.

\begin{table}[htbp]
\caption{The eigenvalues of the $qb\bar{q}\bar{b}$,
$qb\bar{s}\bar{b}$ and $sb\bar{s}\bar{b}$ tetraquark states with
$J^P=0^+, 1^+, 2^+$.} \label{tab:Vcm-eigenvalue-b}
\begin{center}
\begin{tabular}{c| c  | c c c c c c}
\hline \hline configuration  &  $J^P$   & \multicolumn{6}{c}{$V_{CM}$ }  \\
\hline
\multirow{3}{*}{$qb\bar{q}\bar{b}$} & $0^+$ &   $-16.2v$ & $-3.0v$ & $1.8v$ & $5.8v$  \\
\cline{2-8}
                                    & $1^+$ &   $-16.0v$ & $-1.8v$ & $-0.8v$ & $2.0v$ & $5.3v$ & $5.4v$  \\
\cline{2-8}
                                    & $2^+$ &   $0.4v$ & $5.4v$    \\
\hline\hline
\multirow{3}{*}{$qb\bar{s}\bar{b}$} & $0^+$ &   $-13.1v$ & $-2.7v$ & $1.5v$ & $4.7v$  \\
\cline{2-8}
                                  & $1^+$ &   $-12.9v$ & $-1.6v$ & $-0.6v$ & $1.6v$ & $4.31v$ & $4.34v$  \\
\cline{2-8}
                                  & $2^+$ &   $0.5v$ & $4.3v$  \\
\hline\hline
\multirow{3}{*}{$sb\bar{s}\bar{b}$} & $0^+$ &   $-10.7v$ & $-2.3v$ & $1.1v$ & $3.9v$  \\
\cline{2-8}
                                  & $1^+$ &   $-10.4v$ & $-1.3v$ & $-0.5v$ & $1.3v$ & $3.48v$ & $3.51v$  \\
\cline{2-8}
                                  & $2^+$ &   $0.5v$ & $3.5v$  \\
\hline \hline
\end{tabular}
\end{center}
\end{table}

\begin{table}[htbp]
\caption{The masses of the $1^+$ $qb\bar{q}\bar{b}$ tetraquark
states. The parameter $v$ is fixed using the $Z_c$ mass as input. }
\label{tab:v-mass-b}
\begin{center}
\begin{tabular}{c  c  c  c  c  c  c  c  c}
\hline \hline  $v$ &  $-16.0v$ & $-1.8v$ & $-0.8v$ & $2.0v$ & $5.3v$ & $5.4v$ \\
\hline
$v_{-15.9v}^{Z_c(3900)} = 2.6$ & $10292.4$   &  $10329.3$  &  $10331.9$  &  $10339.2$ &  $10347.8$  &  $10348$\\
\hline
$v_{-4.1}^{Z_c(3900)} = 10.2$ & $10170.8$   &  $10315.6$  &  $10325.8$  &  $10354.4$ &  $10388.1$  &  $10389.1$\\
\hline
$v_{-1.5}^{Z_c(3900)} = 28.0$ & $9886$  &  $10283.6$  &  $10311.6$  &  $10390$  &  $10482.4$  &  $10485.2$\\
\hline
$v_{1.7}^{Z_c(4025)} = 48.8$ & $9553.2$  &  $10246.2$  &  $10295$  &  $10431.6$  &  $10592.6$  &  $10597.5$\\
\hline
$v_{5.6}^{Z_c(4025)} = 14.8$ & $10097.2$  &  $10307.4$  &  $10322.2$  &  $10363.6$  &  $10412.4$  &  $10413.9$\\
\hline
$v_{5.8} ^{Z_c(4025)}= 14.3$ & $10105.2$  &  $10308.3$  &  $10322.6$  &  $10362.6$  &  $10409.8$  &  $10411.2$\\
\hline
$v_{5.6}^{Z_c(4200)} = 46.1$ & $9596.4$  &  $10251$  &  $10297.1$  &  $10426.2$  &  $10578.3$  &  $10582.9$\\
\hline
$v_{5.8} ^{Z_c(4200)}= 44.5$ & $9622$  &  $10253.9$  &  $10298.4$  &  $10423$  &  $10569.9$  &  $10574.3$\\
\hline\hline
\end{tabular}
\end{center}
\end{table}

\begin{table}[htbp]
\caption{The eigenvalues and masses of the $qb\bar{q}\bar{b}$,
$qb\bar{s}\bar{b}$ and $sb\bar{s}\bar{b}$ tetraquark states with
$J^P=0^+, 1^+, 2^+$. The parameter $v$ is fixed assuming $Z(4025)$
as the tetraquark state with the eigenvalue $1.7v$. }
\label{tab:mass-s-b}
\begin{center}
\begin{tabular}{c| c  | c c c }
\hline \hline   &     &  $0^+$ & $1^+$ & $2^+$   \\
\hline
\multirow{4}{*}{$qb\bar{q}\bar{b}$} & $V_{CM}$ &   $-16.2v$ & $-16.0v$ & $0.4v$   \\
\cline{2-5}
                                    & $M$(MeV) &   $9543.1$ & $9552.8$ & $10353.5$   \\
\cline{2-5}
                                    & $V_{CM}$ &   $-3.0v$ & $-1.8v$   & $5.4v$   \\
\cline{2-5}
                                    & $M$(MeV) &   $10187.5$ & $10246.1$ & $10597.6$   \\
\hline\hline
\multirow{4}{*}{$qb\bar{s}\bar{b}$} & $V_{CM}$ &   $-13.1v$ & $-12.9v$ & $0.5v$   \\
\cline{2-5}
                                    & $M$(MeV) &   $9799.4$ & $9809.2$ & $10463.4$   \\
\cline{2-5}
                                    & $V_{CM}$ &   $-2.7v$  & $-1.6v$   & $4.3v$   \\
\cline{2-5}
                                    & $M$(MeV) &   $10307.2$ & $10360.9$ & $10648.9$   \\
\hline\hline
\multirow{2}{*}{$sb\bar{s}\bar{b}$} & $V_{CM}$ &   $-10.7v$ & $-10.4v$ & $0.5v$   \\
\cline{2-5}
                                    & $M$(MeV) &   $10021.6$ & $10036.2$ & $10568.4$   \\
\cline{2-5}
                                    & $V_{CM}$ &   $-2.3v$  & $-1.3v$   & $3.5v$   \\
\cline{2-5}
                                    & $M$(MeV) &   $10431.7$ & $10480.5$ & $10714.9$   \\
\hline \hline
\end{tabular}
\end{center}
\end{table}

\section{Decay patterns of hidden-charm tetraquarks}\label{decay}

The eigenvalues of the CM interaction matrix $V_{CM}$ can be used to
derive the mass of tetraquark system, while the eigenvectors of
$V_{CM}$ contain important information on their decay pattern.
Therefore, we carefully investigate the eigenvectors of the
tetraquark systems with the configuration $qc\bar{q}\bar{c}$,
$qc\bar{s}\bar{c}$ and $sc\bar{s}\bar{c}$ and $J^P=0^+, 1^+, 2^+$.
We first list the eigenvalues of $V_{CM}$ for the
$qc\bar{q}\bar{c}$, $qc\bar{s}\bar{c}$ and $sc\bar{s}\bar{c}$
tetraquark configuration in Table \ref{tab:Vcm-eigenvalue}.

\begin{table}[htbp]
\caption{The eigenvalues of $V_{CM}$ for the tetraquark
configuration $qc\bar{q}\bar{c}$, $qc\bar{s}\bar{c}$ and
$sc\bar{s}\bar{c}$ with $J^P=0^+, 1^+, 2^+$.}
\label{tab:Vcm-eigenvalue}
\begin{center}
\begin{tabular}{c| c  | c c c c c c}
\hline \hline configuration  &  $J^P$   & \multicolumn{6}{c}{$V_{CM}$ }  \\
\hline
\multirow{3}{*}{$qc\bar{q}\bar{c}$} & $0^+$ &   $-18.6v$ & $-7.4v$ & $0.8v$ & $8.3v$  \\
\cline{2-8}
                                    & $1^+$ &   $-15.9v$ & $-4.1v$ & $-1.5v$ & $1.7v$ & $5.6v$ & $5.8v$  \\
\cline{2-8}
                                    & $2^+$ &   $2.7v$ & $5.8v$    \\
\hline\hline
\multirow{3}{*}{$qc\bar{s}\bar{c}$} & $0^+$ &   $-15.6v$ & $-6.5v$ & $0.5v$ & $7.1v$  \\
\cline{2-8}
                                  & $1^+$ &   $-12.8v$ & $-3.8v$ & $-1.3v$ & $1.3v$ & $4.6v$ & $4.8v$  \\
\cline{2-8}
                                  & $2^+$ &   $2.5v$ & $4.7v$  \\
\hline\hline
\multirow{3}{*}{$sc\bar{s}\bar{c}$} & $0^+$ &   $-13.1v$ & $-5.7v$ & $0.3v$ & $6.1v$  \\
\cline{2-8}
                                  & $1^+$ &   $-10.3v$ & $-3.2v$ & $-1.3v$ & $1.0v$ & $3.8v$ & $3.9v$  \\
\cline{2-8}
                                  & $2^+$ &   $2.3v$ & $3.9v$  \\
\hline \hline
\end{tabular}
\end{center}
\end{table}

For the $J^P=0^+, 1^+$ case, we only list the eigenvectors with the
negative eigenvalues. When we present the eigenvectors using the
diquark representation $q\bar{q} \otimes q\bar{q}$, we omit the $N$
in the diquark representation $|D_6, D_{3c}, S, N \rangle$ for
brevity since $N=2$. We present the expressions of the eigenvectors
for the $qc\bar{q}\bar{c}$, $qc\bar{s}\bar{c}$, and
$sc\bar{s}\bar{c}$ tetraquark systems in Tables
\ref{tab:Vqcqc-eigenvectors1}- \ref{tab:Vqcqc-eigenvectors2},
\ref{tab:Vqcsc-eigenvectors1}-\ref{tab:Vqcsc-eigenvectors2}, and
\ref{tab:Vscsc-eigenvectors1}-\ref{tab:Vscsc-eigenvectors2}
respectively.

We notice that $J/\psi$ and $\eta_c$ can also be expressed as the
$SU(6)_{cs}$ $c\bar{c}$ state $|35, 1_c, 1\rangle$ and $|1, 1_c,
0\rangle$. Similarly, $D^*$ and $D$ can be treated as the
$SU(6)_{cs}$ $c\bar{u}$ state $|35, 1_c, 1\rangle$ and $|1, 1_c,
0\rangle$. Therefore, we can identify the decay patterns of the
tetraquark states from the expression of their CM interaction
eigenvectors. The branching fraction of each decay mode is
proportional to the square of the coefficient of the corresponding
component in the eigenvectors if we ignore the phase space
difference. From the very beginning, we want to emphasize the
following point: so long as the phase space allows, the $\psi' \pi$
decay mode is also allowed if $J/\psi\pi$ is one of the allowed
decay modes.

For the $J^P=0^+$ state, the lowest state corresponds to the
eigenvalue $-18.6v$. From Table \ref{tab:Vqcqc-eigenvectors0}, its
dominant decay mode is $\eta_c \pi$. The $\bar{D}D$ mode is also
important. The $\bar{D}^*D^*$ mode is suppressed by a factor of
eight if we compare the coefficients of the $\bar{D}D$ and
$\bar{D}^*D^*$ components only. In fact, the $\bar{D}^*D^*$ mode is
further suppressed by phase space.

The $J^P=0^+$ state with the eigenvalue $-7.4v$ also decays into
$\eta_c \pi$, $\bar{D}D$ and $\bar{D}^*D^*$. However, $\bar{D}D$
becomes its dominant decay mode. The $\bar{D}^*D^*$ mode is also
severely suppressed.

From Table \ref{tab:Vqcqc-eigenvectors0}, the $J^P=2^+$ state with
the eigenvalue $5.8v$ mainly decays into $J/\psi \rho$ while its
$\bar{D}^*D^*$ mode is suppressed. In contrast, the $J^P=2^+$ state with
the eigenvalue $2.7v$ decays into $\bar{D}^*D^*$ only. Its $J/\psi \rho$
mode is forbidden.

Experimentally several charged axial vector hidden-charm states were
reported in different decay channels. All the four charged axial
vector states $Z_c(3900)$, $Z_c(4025)$, $Z_c(4200)$ and
$Z_c(4485)$ were observed in the $J/\psi \pi$ channel.
$Z_c(4485)$ was also observed in the $\psi' \pi$ mode. The
dominant decay mode of $Z_c(3900)$ and $Z_c(4025)$ is $\bar{D}D^*$ and
$\bar{D}^*D^*$ respectively. Up to now, the dominant decay mode of
$Z_c(4200)$ and $Z_c(4485)$ has not been established yet.
Moreover, $Z_c(4025)$ does not decay into $\bar{D}D^*$.

It's very interesting to investigate the decay patterns of the low
lying tetraquark states and compare their typical decay modes with
the available experimental data. From Table
\ref{tab:Vqcqc-eigenvectors1}, the lowest axial vector
$qc\bar{q}\bar{c}$ tetraquark state corresponds to the eigenvalue
$-15.9v$. Its dominant decay mode is $J/\psi \pi$. The $\bar{D}^*D$
mode is suppressed by a factor of sixteen if we compare the
coefficients of the $J/\psi \pi$ and $D^*\bar{D}$ components and
ignore the phase space difference. The $\bar{D}^*D^*$ mode is
further suppressed roughly by a factor of two compared with the
$D^*\bar{D}$ mode. Considering the decay phase space, the
$D^*\bar{D}$ and $\bar{D}^*D^*$ modes are further suppressed. This
state mainly decay into $J/\psi \pi$, which is in strong contrast
with the fact that the dominant decay mode of $Z_c(3900)$ and
$Z_c(4025)$ is $\bar{D}D^*$ and $\bar{D}^*D^*$ respectively. In
other words, neither $Z_c(3900)$ nor $Z_c(4025)$ is a good candidate
of this lowest lying axial vector tetraquark state. On the other
hand, either $Z_c(4200)$ or $Z_c(4485)$ could be a candidate of this
tetraquark state. In fact, $Z_c(4200)$ is a very promising
tetraquark candidate.

The second axial vector tetraquark state with the eigenvalue $-4.1v$
decays into $D^*\bar{D}$ only. It's quite particular that this state
neither decays into $J/\psi \pi$ nor into $\bar{D}^*D^*$ even phase space
allows. Since all the four charged $Z_c$ states decay into the
$J/\psi \pi$ mode, none of them is the candidate of this tetraquark
state.

The third $J^P=1^+$ state corresponds to the eigenvalue $-1.5v$,
which decays into $J/\psi \pi$, $\eta_c \rho$, $D^*\bar{D}$ and
$\bar{D}^*D^*$. Its dominant decay mode is $D^*\bar{D}$. For
comparison, both the $\bar{D}^*D^*$ and $\eta_c \rho$ modes are
suppressed roughly by a factor of eight if we ignore the phase space
difference. In contrast, the $J/\psi \pi$ mode is strongly
suppressed. If we ignore the phase space difference, the suppression
factor is roughly 25 compared with the dominant $D^*\bar{D}$ mode.
Based on the current experimental information, all the three
$Z_c(3900)$, $Z_c(4200)$ and $Z_c(4485)$ can be assigned as this
third tetrquark state with the eigenvalue $-1.5v$. Especially, the
characteristic decay pattern of this third axial vector tetraquark
state matches well with that of $Z_c(3900)$. With such an
assignment, we would expect two more axial vector tetraquark states
with the eigenvalue $-15.9v$ and $-4.1v$ which are very close to (or
even below) the open charm threshold and lie below $Z_c(3900)$.
Their decay patterns are listed in the previous paragraphs.

\begin{table}[htbp]
\caption{The eigenvectors of $V_{CM}$ for the $qc\bar{q}\bar{c}$
tetraquark states with $J^P=1^+$. } \label{tab:Vqcqc-eigenvectors1}
\begin{center}
\begin{tabular}{c  |c  | c  }
\hline \hline     $V_{CM}$   &  $q\bar{q} \otimes c\bar{c}$  &  $c\bar{q} \otimes q\bar{c}$ \\
\hline \multirow{6}{*}{$-15.9v$}
& $+0.99 |1, 1_c, 0 \rangle \otimes |35, 1_c, 1\rangle$   & $+0.24 |1, 1_c, 0 \rangle \otimes |35, 1_c, 1\rangle $ \\
& $+0.14 |35, 8_c, 1 \rangle \otimes |35, 8_c, 1 \rangle$ & $+0.44 |35, 8_c, 0 \rangle \otimes |35, 8_c, 1\rangle$  \\
&                                                        & $+0.24 |35, 1_c, 1 \rangle \otimes |1, 1_c, 0\rangle$  \\
&                                                        & $+0.44 |35, 8_c, 1 \rangle \otimes |35, 8_c, 0\rangle$  \\
&                                                        & $-0.69 |35, 8_c, 1 \rangle \otimes |35, 8_c, 0\rangle$  \\
&                                                        & $-0.15 |35, 1_c, 1 \rangle \otimes |35, 1_c, 1\rangle $ \\
\cline{1-3} \multirow{4}{*}{$-4.1v$}
& $+|35, 8_c, 1 \rangle \otimes |35, 8_c, 1 \rangle$     & $+0.67 |1, 1_c, 0 \rangle \otimes |35, 1_c, 1\rangle $ \\
&                                                        & $-0.24 |35, 8_c, 0 \rangle \otimes |35, 8_c, 1\rangle$  \\
&                                                        & $-0.67 |35, 1_c, 1 \rangle \otimes |1, 1_c, 0\rangle$  \\
&                                                        & $+0.24 |35, 8_c, 1 \rangle \otimes |35, 8_c, 0\rangle$  \\
\cline{1-3} \multirow{6}{*}{$-1.5v$}
&   $+0.12 |1, 1_c, 0 \rangle \otimes |35, 1_c, 1\rangle$ & $-0.65 |1, 1_c, 0 \rangle \otimes |35, 1_c, 1\rangle $\\
&  $-0.56 |35, 8_c, 0 \rangle \otimes |35, 8_c, 1\rangle$ & $+0.18 |35, 8_c, 0 \rangle \otimes |35, 8_c, 1\rangle $  \\
&  $-0.22 |35, 1_c, 1 \rangle \otimes |1, 1_c, 0 \rangle$ & $-0.65 |35, 1_c, 1 \rangle \otimes |1, 1_c, 0\rangle $ \\
&  $-0.79 |35, 8_c, 1\rangle \otimes |35, 8_c, 0 \rangle$ & $+0.18 |35, 8_c, 1 \rangle \otimes |35, 8_c, 0\rangle $ \\
&                                                         & $-0.17 |35, 8_c, 1 \rangle \otimes |35, 8_c, 1\rangle$  \\
&                                                         & $-0.23 |35, 1_c, 1 \rangle \otimes |35, 1_c, 1\rangle$  \\
\hline \hline
\end{tabular}
\end{center}
\end{table}

\begin{table}[htbp]
\caption{The eigenvectors of $V_{CM}$ for the $qc\bar{q}\bar{c}$
tetraquark states with $J^P=0^+$. } \label{tab:Vqcqc-eigenvectors0}
\begin{center}
\begin{tabular}{c  |c  | c  }
\hline \hline     $V_{CM}$   &  $q\bar{q} \otimes c\bar{c}$  &  $c\bar{q} \otimes q\bar{c}$ \\
\hline \multirow{4}{*}{$-18.6v$}
& $+0.94 |1, 1_c, 0 \rangle \otimes |1, 1_c, 0\rangle$   & $+0.45|1, 1_c, 0 \rangle \otimes |1, 1_c, 0\rangle $ \\
& $+0.33 |35, 8_c, 1 \rangle \otimes |35, 8_c, 1 \rangle$ & $-0.16 |35, 1_c, 1 \rangle \otimes |35, 1_c, 1\rangle$  \\
&                                                        & $+0.35 |35, 8_c, 0 \rangle \otimes |35, 8_c, 0\rangle$  \\
&                                                        & $-0.80 |35, 8_c, 1 \rangle \otimes |35, 8_c, 1\rangle$  \\
\cline{1-3} \multirow{4}{*}{$-7.4v$}
& $+0.33 |1, 1_c, 0 \rangle \otimes |1, 1_c, 0 \rangle$     & $-0.85 |1, 1_c, 0 \rangle \otimes |1, 1_c, 0\rangle $ \\
& $-0.34 |35, 8_c, 0 \rangle \otimes |35, 8_c, 0\rangle$   & $-0.25 |35, 1_c, 1 \rangle \otimes |35, 1_c, 1\rangle$  \\
& $+0.88 |35, 8_c, 1 \rangle \otimes |35, 8_c, 1\rangle$   & $-0.39 |35, 8_c, 0 \rangle \otimes |35, 8_c, 0\rangle$  \\
&                                                        & $-0.27 |35, 8_c, 1 \rangle \otimes |35, 8_c, 1\rangle$  \\
\hline \hline
\end{tabular}
\end{center}
\end{table}

\begin{table}[htbp]
\caption{The eigenvectors of $V_{CM}$ for the $qc\bar{q}\bar{c}$
tetraquark states with $J^P=2^+$.  } \label{tab:Vqcqc-eigenvectors2}
\begin{center}
\begin{tabular}{c  |c  | c  }
\hline \hline     $V_{CM}$   &  $q\bar{q} \otimes c\bar{c}$  &  $c\bar{q} \otimes q\bar{c}$ \\
\hline \multirow{2}{*}{$5.8v$}
& $+|35, 1_c, 1 \rangle \otimes |35, 1_c, 1\rangle$   & $+0.94|35, 8_c, 1 \rangle \otimes |35, 8_c, 1\rangle $ \\
&                                                                                & $+0.33 |35, 1_c, 1 \rangle \otimes |35, 1_c, 1\rangle$  \\
\cline{1-3} \multirow{2}{*}{$2.7v$}
& $-|35, 8_c, 1 \rangle \otimes |35, 8_c, 1 \rangle$     & $+0.33 |35, 8_c, 1 \rangle \otimes |35, 8_c, 1\rangle $ \\
&                                                                                  & $-0.94 |35, 1_c, 1 \rangle \otimes |35, 1_c, 1\rangle$  \\
\hline \hline
\end{tabular}
\end{center}
\end{table}

\begin{table}[htbp]
\caption{The eigenvectors of $V_{CM}$ for the $qc\bar{s}\bar{c}$
tetraquark states with $J^P=1^+$. } \label{tab:Vqcsc-eigenvectors1}
\begin{center}
\begin{tabular}{c  |c  | c  }
\hline \hline     $V_{CM}$   &  $q\bar{s} \otimes c\bar{c}$  &  $c\bar{s} \otimes q\bar{c}$ \\
\hline \multirow{6}{*}{$-12.8v$}
& $+0.99 |1, 1_c, 0 \rangle \otimes |35, 1_c, 1\rangle$   & $+0.24 |1, 1_c, 0 \rangle \otimes |35, 1_c, 1\rangle $ \\
& $+0.16 |35, 8_c, 1 \rangle \otimes |35, 8_c, 0 \rangle$ & $+0.43 |35, 8_c, 0 \rangle \otimes |35, 8_c, 1\rangle$  \\
&                                                        & $+0.25 |35, 1_c, 1 \rangle \otimes |1, 1_c, 0\rangle$  \\
&                                                        & $+0.43 |35, 8_c, 1 \rangle \otimes |35, 8_c, 0\rangle$  \\
&                                                        & $-0.69 |35, 8_c, 1 \rangle \otimes |35, 8_c, 0\rangle$  \\
&                                                        & $-0.14 |35, 1_c, 1 \rangle \otimes |35, 1_c, 1\rangle $ \\
\cline{1-3} \multirow{4}{*}{$-3.8v$}
& $-0.16 |35, 8_c, 0 \rangle \otimes |35, 8_c, 1\rangle$     & $+0.5 |1, 1_c, 0 \rangle \otimes |35, 1_c, 1\rangle $ \\
& $-0.14 |35, 8_c, 1 \rangle \otimes |35, 8_c, 0\rangle$       & $-0.2 |35, 8_c, 0 \rangle \otimes |35, 8_c, 1\rangle$  \\
& $-0.97 |35, 8_c, 1 \rangle \otimes |35, 8_c, 1\rangle$         & $-0.8 |35, 1_c, 1 \rangle \otimes |1, 1_c, 0\rangle$  \\
&                                                        & $+0.25 |35, 8_c, 1 \rangle \otimes |35, 8_c, 0\rangle$  \\
\cline{1-3} \multirow{6}{*}{$-1.3v$}
&  $+0.13 |1, 1_c, 0 \rangle \otimes |35, 1_c, 1\rangle$ & $-0.78 |1, 1_c, 0 \rangle \otimes |35, 1_c, 1\rangle $\\
&  $-0.57 |35, 8_c, 0 \rangle \otimes |35, 8_c, 1\rangle$ & $+0.24 |35, 8_c, 0 \rangle \otimes |35, 8_c, 1\rangle $  \\
&  $-0.24 |35, 1_c, 1 \rangle \otimes |1, 1_c, 0 \rangle$ & $-0.49 |35, 1_c, 1 \rangle \otimes |1, 1_c, 0\rangle $ \\
&  $-0.74 |35, 8_c, 1\rangle \otimes |35, 8_c, 0 \rangle$ & $-0.09 |35, 8_c, 1 \rangle \otimes |35, 8_c, 0\rangle $ \\
&   $+0.22 |35, 8_c, 1 \rangle \otimes |35, 8_c, 1\rangle$    & $-0.21 |35, 8_c, 1 \rangle \otimes |35, 8_c, 1\rangle$  \\
&                                                         & $-0.2 |35, 1_c, 1 \rangle \otimes |35, 1_c, 1\rangle$  \\
\hline \hline
\end{tabular}
\end{center}
\end{table}

\begin{table}[htbp]
\caption{The eigenvectors of $V_{CM}$ for the $qc\bar{s}\bar{c}$
tetraquark states with $J^P=0^+$. } \label{tab:Vqcsc-eigenvectors0}
\begin{center}
\begin{tabular}{c  |c  | c  }
\hline \hline     $V_{CM}$   &  $q\bar{s} \otimes c\bar{c}$  &  $c\bar{s} \otimes q\bar{c}$ \\
\hline \multirow{4}{*}{$-15.6v$}
& $+0.92 |1, 1_c, 0 \rangle \otimes |1, 1_c, 0\rangle$   & $+0.5 |1, 1_c, 0 \rangle \otimes |1, 1_c, 0\rangle $ \\
& $-0.38 |35, 8_c, 1 \rangle \otimes |35, 8_c, 1 \rangle$ & $-0.14 |35, 1_c, 1 \rangle \otimes |35, 1_c, 1\rangle$  \\
&                                                        & $+0.33 |35, 8_c, 0 \rangle \otimes |35, 8_c, 0\rangle$  \\
&                                                        & $-0.79 |35, 8_c, 1 \rangle \otimes |35, 8_c, 1\rangle$  \\
\cline{1-3} \multirow{4}{*}{$-6.5v$}
& $+0.37 |1, 1_c, 0 \rangle \otimes |1, 1_c, 0 \rangle$     & $-0.83 |1, 1_c, 0 \rangle \otimes |1, 1_c, 0\rangle $ \\
& $-0.35 |35, 8_c, 0 \rangle \otimes |35, 8_c, 0\rangle$   & $-0.24 |35, 1_c, 1 \rangle \otimes |35, 1_c, 1\rangle$  \\
& $+0.85 |35, 8_c, 1 \rangle \otimes |35, 8_c, 1\rangle$   & $+0.4 |35, 8_c, 0 \rangle \otimes |35, 8_c, 0\rangle$  \\
&                                                        & $-0.31 |35, 8_c, 1 \rangle \otimes |35, 8_c, 1\rangle$  \\
\hline \hline
\end{tabular}
\end{center}
\end{table}

\begin{table}[htbp]
\caption{The eigenvectors of $V_{CM}$ for the $qc\bar{s}\bar{c}$
tetraquark states with $J^P=2^+$. } \label{tab:Vqcsc-eigenvectors2}
\begin{center}
\begin{tabular}{c  |c  | c  }
\hline \hline     $V_{CM}$   &  $q\bar{s} \otimes c\bar{c}$  &  $c\bar{s} \otimes q\bar{c}$ \\
\hline \multirow{2}{*}{$4.7v$}
& $+|35, 1_c, 1 \rangle \otimes |35, 1_c, 1\rangle$   & $+0.94|35, 8_c, 1 \rangle \otimes |35, 8_c, 1\rangle $ \\
&                                                                                & $+0.33 |35, 1_c, 1 \rangle \otimes |35, 1_c, 1\rangle$  \\
\cline{1-3} \multirow{2}{*}{$2.5v$}
& $+|35, 8_c, 1 \rangle \otimes |35, 8_c, 1 \rangle$     & $+0.33 |35, 8_c, 1 \rangle \otimes |35, 8_c, 1\rangle $ \\
&                                                                                  & $-0.94 |35, 1_c, 1 \rangle \otimes |35, 1_c, 1\rangle$  \\
\hline \hline
\end{tabular}
\end{center}
\end{table}

\begin{table}[htbp]
\caption{The eigenvectors of $V_{CM}$ for the $sc\bar{s}\bar{c}$
tetraquark states with $J^P=1^+$.} \label{tab:Vscsc-eigenvectors1}
\begin{center}
\begin{tabular}{c  |c  | c  }
\hline \hline     $V_{CM}$   &  $s\bar{s} \otimes c\bar{c}$  &  $c\bar{s} \otimes s\bar{c}$ \\
\hline \multirow{6}{*}{$-10.3v$}
& $+0.98 |1, 1_c, 0 \rangle \otimes |35, 1_c, 1\rangle$   & $+0.25 |1, 1_c, 0 \rangle \otimes |35, 1_c, 1\rangle $ \\
& $+0.17 |35, 8_c, 1 \rangle \otimes |35, 8_c, 0 \rangle$ & $+0.43 |35, 8_c, 0 \rangle \otimes |35, 8_c, 1\rangle$  \\
&                                                        & $+0.26 |35, 1_c, 1 \rangle \otimes |1, 1_c, 0\rangle$  \\
&                                                        & $+0.43 |35, 8_c, 1 \rangle \otimes |35, 8_c, 0\rangle$  \\
&                                                        & $-0.69 |35, 8_c, 1 \rangle \otimes |35, 8_c, 0\rangle$  \\
&                                                        & $-0.13 |35, 1_c, 1 \rangle \otimes |35, 1_c, 1\rangle $ \\
\cline{1-3} \multirow{4}{*}{$-3.2v$}
& $+|35, 8_c, 1 \rangle \otimes |35, 8_c, 1\rangle$     & $+0.67 |1, 1_c, 0 \rangle \otimes |35, 1_c, 1\rangle $ \\
&                                                                                 & $-0.24 |35, 8_c, 0 \rangle \otimes |35, 8_c, 1\rangle$  \\
&                                                                                 & $-0.67 |35, 1_c, 1 \rangle \otimes |1, 1_c, 0\rangle$  \\
&                                                        & $+0.24 |35, 8_c, 1 \rangle \otimes |35, 8_c, 0\rangle$  \\
\cline{1-3} \multirow{6}{*}{$-1.3v$}
&  $+0.14 |1, 1_c, 0 \rangle \otimes |35, 1_c, 1\rangle$ & $-0.65 |1, 1_c, 0 \rangle \otimes |35, 1_c, 1\rangle $\\
&  $-0.63 |35, 8_c, 0 \rangle \otimes |35, 8_c, 1\rangle$ & $+0.14 |35, 8_c, 0 \rangle \otimes |35, 8_c, 1\rangle $  \\
&  $-0.31 |35, 1_c, 1 \rangle \otimes |1, 1_c, 0 \rangle$ & $-0.65 |35, 1_c, 1 \rangle \otimes |1, 1_c, 0\rangle $ \\
&  $-0.7 |35, 8_c, 1\rangle \otimes |35, 8_c, 0 \rangle$ & $-0.14 |35, 8_c, 1 \rangle \otimes |35, 8_c, 0\rangle $ \\
&                                                                                    & $-0.29 |35, 8_c, 1 \rangle \otimes |35, 8_c, 1\rangle$  \\
&                                                                                   & $-0.15 |35, 1_c, 1 \rangle \otimes |35, 1_c, 1\rangle$  \\
\hline \hline
\end{tabular}
\end{center}
\end{table}

\begin{table}[htbp]
\caption{The eigenvectors of $V_{CM}$ for the $sc\bar{s}\bar{c}$
tetraquark states with $J^P=0^+$. } \label{tab:Vscsc-eigenvectors0}
\begin{center}
\begin{tabular}{c  |c  | c  }
\hline \hline     $V_{CM}$   &  $s\bar{s} \otimes c\bar{c}$  &  $c\bar{s} \otimes s\bar{c}$ \\
\hline \multirow{4}{*}{$-13.1v$}
& $+0.91 |1, 1_c, 0 \rangle \otimes |1, 1_c, 0\rangle$   & $+0.53 |1, 1_c, 0 \rangle \otimes |1, 1_c, 0\rangle $ \\
& $-0.41 |35, 8_c, 1 \rangle \otimes |35, 8_c, 1 \rangle$ & $-0.13 |35, 1_c, 1 \rangle \otimes |35, 1_c, 1\rangle$  \\
&                                                        & $+0.31 |35, 8_c, 0 \rangle \otimes |35, 8_c, 0\rangle$  \\
&                                                        & $-0.78 |35, 8_c, 1 \rangle \otimes |35, 8_c, 1\rangle$  \\
\cline{1-3} \multirow{4}{*}{$-5.6v$}
& $+0.41 |1, 1_c, 0 \rangle \otimes |1, 1_c, 0 \rangle$     & $-0.81 |1, 1_c, 0 \rangle \otimes |1, 1_c, 0\rangle $ \\
& $-0.36 |35, 8_c, 0 \rangle \otimes |35, 8_c, 0\rangle$   & $-0.24 |35, 1_c, 1 \rangle \otimes |35, 1_c, 1\rangle$  \\
& $+0.83 |35, 8_c, 1 \rangle \otimes |35, 8_c, 1\rangle$   & $+0.4 |35, 8_c, 0 \rangle \otimes |35, 8_c, 0\rangle$  \\
&  $+0.11 |35, 1_c, 1 \rangle \otimes |35, 1_c, 1\rangle$   & $-0.35 |35, 8_c, 1 \rangle \otimes |35, 8_c, 1\rangle$  \\
\hline \hline
\end{tabular}
\end{center}
\end{table}

\begin{table}[htbp]
\caption{The eigenvectors of $V_{CM}$ for the $sc\bar{s}\bar{c}$
tetraquark states with $J^P=2^+$. } \label{tab:Vscsc-eigenvectors2}
\begin{center}
\begin{tabular}{c  |c  | c  }
\hline \hline     $V_{CM}$   &  $s\bar{s} \otimes c\bar{c}$  &  $c\bar{s} \otimes s\bar{c}$ \\
\hline \multirow{2}{*}{$3.9v$}
& +$|35, 1_c, 1 \rangle \otimes |35, 1_c, 1\rangle$   & $+0.94|35, 8_c, 1 \rangle \otimes |35, 8_c, 1\rangle $ \\
&                                                                                & $+0.33 |35, 1_c, 1 \rangle \otimes |35, 1_c, 1\rangle$  \\
\cline{1-3} \multirow{2}{*}{$2.3v$}
& $+|35, 8_c, 1 \rangle \otimes |35, 8_c, 1 \rangle$     & $+0.33 |35, 8_c, 1 \rangle \otimes |35, 8_c, 1\rangle $ \\
&                                                                                  & $-0.94 |35, 1_c, 1 \rangle \otimes |35, 1_c, 1\rangle$  \\
\hline \hline
\end{tabular}
\end{center}
\end{table}

\section{Summary}\label{summary}

Within the framework of the color-magnetic interaction, we have
systematically considered the mass spectrum of the hidden-charm and
hidden-bottom tetraquark states with the configurations
$qc\bar{q}\bar{c}$, $qc\bar{s}\bar{c}$, $sc\bar{s}\bar{c}$,
$qb\bar{q}\bar{b}$, $qb\bar{s}\bar{b}$, $sb\bar{s}\bar{b}$ and
$J^P=1^+, 0^+, 2^+$.

Experimentally several charged axial-vector hidden-charm states were
reported. We have adopted two schemes to fix the parameter $v$ and
extracted the tetraquark spectrum. We first tried to assume one of
$Z_c$ states is a tetraquark state and use its mass as input to
determine $v$ and the masses of the other tetraquark states. We
notice that it is impossible to accommodate all the three charged
states $Z_c(3900)$, $Z_c(4025)$ and $Z_c(4200)$ within the axial
vector tetraquark spectrum simultaneously. Then we tried to use the
mass splitting between two $Z_c$ states as input. With the second
scheme we could accommodate all three charged states $Z_c(3900)$,
$Z_c(4025)$ and $Z_c(4200)$ as the axial vector tetraquark
candidates simultaneously. However, the resulting mass of the lowest
axial vector tetraquark state is always too low and unrealistic. We
have to conclude that not all these three states are tetraquark
candidates. Instead of being a tetraquark candidate, at least one or
two of these states is probably a molecular state or some other
structure.

Moreover, the eigenvectors of the chromomagnetic interaction
contains valuable information of the decay pattern of the tetraquark
states. For example, the dominant decay mode of the lowest axial
vector $qc\bar{q}\bar{c}$ tetraquark state is $J/\psi \pi$. Its
$D^*\bar{D}$ and $\bar{D}^*D^*$ modes are strongly suppressed. Recall that the
dominant decay mode of $Z_c(3900)$ and $Z_c(4025)$ is $\bar{D}D^*$ and
$\bar{D}^*D^*$ respectively. We tend to conclude that neither $Z_c(3900)$
nor $Z_c(4025)$ is a good candidate of the lowest lying axial
vector tetraquark state. In fact, $Z_c(3900)$ and $Z_c(4025)$ is
close to the $\bar{D}D^*$ and $\bar{D}^*D^*$ mass threshold. They are good
molecular candidates. Their mass and decay pattern agree with the
naive expectation within the molecular picture.

On the other hand, the charmonium-like charged state $Z_c(4200)$ is
observed in the $J/\psi \pi$ channel with significance $8.2\sigma$.
Its mass is far away from the mass threshold of two S-wave heavy
mesons. In fact, the axial vector hidden-charm tetraquark state was
predicted to lie around 4.2 GeV several years ago \cite{chenwei}. As
expected as a tetraquark candidate, $Z_c(4200)$ is very broad with a
width around 370 MeV. All the available experimental information
indicates that $Z_c(4200)$ is a very promising candidate of the
lowest axial vector hidden-charm tetraquark state. Future
experimental investigations of this state will be very desirable.

\section*{Acknowledgment}
We thank Li-Ping Sun for useful discussions. This project is
supported by the National Natural Science Foundation of China under
Grant No. 11261130311.

\section*{Appendix}

The $SU(6)_{cs}$ eigenstates of the $0^+$ tetraquark in the $Q
\otimes \bar{Q}$ form are:
\begin{eqnarray}
| 1, 1_c, 0, 4 \rangle &=& \sqrt{\frac{6}{7}}| 21, 6_c, 1, 2 \rangle \otimes |
\bar{21}, \bar{6}_c, 1, 2 \rangle \nonumber\\ &+&
\sqrt{\frac{1}{7}}| 21, \bar{3}_c, 0, 2 \rangle \otimes |
\bar{21}, 3_c, 0, 2 \rangle
\end{eqnarray}
\begin{eqnarray}
| 405, 1_c, 0, 4 \rangle &=& \sqrt{\frac{1}{7}}| 21, 6_c, 1, 2 \rangle \otimes |
\bar{21}, \bar{6}_c, 1, 2 \rangle \nonumber\\ &-&
\sqrt{\frac{6}{7}}| 21, \bar{3}_c, 0, 2 \rangle \otimes |
\bar{21}, 3_c, 0, 2 \rangle
\end{eqnarray}
\begin{eqnarray}
| 1, 1_c, 0, 4 \rangle &=& \sqrt{\frac{3}{5}} | 15, \bar{3}_c, 1, 2
\rangle \otimes | \bar{15}, 3_c, 1, 2 \rangle \nonumber\\ &+&
\sqrt{\frac{2}{5}} | 15, 6_c, 0, 2 \rangle \otimes | \bar{15},
\bar{6}_c, 0, 2 \rangle
\end{eqnarray}
\begin{eqnarray}
| 189, 1_c, 0, 4 \rangle &=& \sqrt{\frac{2}{5}} | 15, \bar{3}_c, 1,
2 \rangle \otimes | \bar{15}, 3_c, 1, 2 \rangle \nonumber\\ &-&
\sqrt{\frac{3}{5}} | 15, 6_c, 0, 2 \rangle \otimes | \bar{15}, \bar{6}_c, 0, 2 \rangle
\end{eqnarray}

The $SU(6)_{cs}$ eigenstates of the $0^+$ tetraquark in the $q_1
\bar{q}_3 \otimes q_2 \bar{q}_4 $ form are:
\begin{eqnarray}
| 1, 1_c, 0, 4 \rangle &=& \frac{\sqrt{21}}{6} | q_1 \bar{q}_3 1,
1_c, 0, 2 \rangle \otimes | q_2 \bar{q}_4 1, 1_c, 0, 2 \rangle
\nonumber\\ &+& \frac{\sqrt{7}}{14} | q_1 \bar{q}_3 35, 1_c, 1, 2
\rangle \otimes | q_2 \bar{q}_4 35, 1_c, 1, 2 \rangle \nonumber\\
&+& \frac{\sqrt{42}}{21} | q_1 \bar{q}_3 35, 8_c, 0, 2 \rangle \otimes
| q_2 \bar{q}_4 35, 8_c, 0, 2 \rangle \nonumber\\ &-&
\frac{\sqrt{14}}{7} | q_1 \bar{q}_3 35, 8_c, 1, 2 \rangle \otimes |
q_2 \bar{q}_4 35, 8_c, 1, 2 \rangle
\end{eqnarray}
\begin{eqnarray}
| 405, 1_c, 0, 4 \rangle &=& \frac{2\sqrt{42}}{21} | q_1 \bar{q}_3 35,
1_c, 1, 2 \rangle \otimes | q_2 \bar{q}_4 35, 1_c, 1, 2 \rangle
\nonumber\\ &+& \frac{3\sqrt{7}}{14} | q_1 \bar{q}_3 35, 8_c, 0, 2
\rangle \otimes | q_2 \bar{q}_4 35, 8_c, 0, 2 \rangle \nonumber\\
&+& \frac{5\sqrt{21}}{42} | q_1 \bar{q}_3 35, 8_c, 1, 2 \rangle \otimes
| q_2 \bar{q}_4 35, 8_c, 1, 2 \rangle
\end{eqnarray}
\begin{eqnarray}
| 1, 1_c, 0, 4 \rangle &=& \frac{15}{6} | q_1 \bar{q}_3 1, 1_c, 0, 2
\rangle \otimes | q_2 \bar{q}_4 1, 1_c, 0, 2 \rangle \nonumber\\
&+& \frac{5}{10} | q_1 \bar{q}_3 35, 1_c, 1, 2 \rangle \otimes | q_2
\bar{q}_4 35, 1_c, 1, 2 \rangle \nonumber\\ &-& \frac{\sqrt{30}}{15} | q_1
\bar{q}_3 35, 8_c, 0, 2 \rangle \otimes | q_2 \bar{q}_4 35, 8_c, 0,
2 \rangle \nonumber\\ &+& \frac{\sqrt{10}}{5} | q_1 \bar{q}_3 35,
8_c, 1, 2 \rangle \otimes | q_2 \bar{q}_4 35, 8_c, 1, 2 \rangle
\end{eqnarray}
\begin{eqnarray}
| 189, 1_c, 0, 4 \rangle &=& -\frac{2\sqrt{30}}{15} | q_1 \bar{q}_3 35, 1_c,
1, 2 \rangle \otimes | q_2 \bar{q}_4 35, 1_c, 1, 2 \rangle
\nonumber\\ &-& \frac{3\sqrt{5}}{10} | q_1 \bar{q}_3 35, 8_c, 0, 2 \rangle
\otimes | q_2 \bar{q}_4 35, 8_c, 0, 2 \rangle \nonumber\\ &-&
\frac{\sqrt{15}}{30} | q_1 \bar{q}_3 35, 8_c, 1, 2 \rangle \otimes |
q_2 \bar{q}_4 35, 8_c, 1, 2 \rangle
\end{eqnarray}

According to the $SU(3)_c$ and $SU(2)_s$ symmetry, the $SU(6)_{cs}$
eigenstates of the $q_2 \bar{q}_3 \otimes q_1 \bar{q}_4$ form are
the same as those of the $q_1 \bar{q}_3 \otimes q_2 \bar{q}_4$ form
in the first two tetraquark states, while there appears an extra
minus sign in the last two tetraquark states.

The $SU(6)_{cs}$ eigenstates of the $2^+$ tetraquark in the $Q
\otimes \bar{Q}$ form are:
\begin{eqnarray}
| 405, 1_c, 0, 4 \rangle = | 21, 6_c, 1, 2 \rangle \otimes |
\bar{21}, \bar{6}_c, 1, 2 \rangle
\end{eqnarray}
\begin{eqnarray}
| 189, 1_c, 2, 4 \rangle =  | 15, \bar{3}_c, 1, 2 \rangle \otimes | \bar{15}, 3_c, 1, 2 \rangle
\end{eqnarray}
The $SU(6)_{cs}$ eigenstates of the $2^+$ tetraquark in the $q_1
\bar{q}_3 \otimes q_2 \bar{q}_4 $ form are:
\begin{eqnarray}
| 405, 1_c, 2, 4 \rangle &=& \frac{1}{3} | q_1 \bar{q}_3 35, 8_c, 1, 2
\rangle \otimes | q_2 \bar{q}_4 35, 8_c, 1, 2 \rangle \nonumber\\
&+& \frac{2}{3} | q_1 \bar{q}_3 35, 1_c, 1, 2 \rangle \otimes | q_2
\bar{q}_4 35, 1_c, 1, 2 \rangle
\end{eqnarray}
\begin{eqnarray}
| 189, 1_c, 2, 4 \rangle &=& -\frac{\sqrt{2}}{3} | q_1 \bar{q}_3 35, 8_c,
1, 2 \rangle \otimes | q_2 \bar{q}_4 35, 8_c, 1, 2 \rangle
\nonumber\\ &-& \frac{3\sqrt{1}}{3} | q_1 \bar{q}_3 35, 1_c, 1, 2 \rangle
\otimes | q_2 \bar{q}_4 35, 1_c, 1, 2 \rangle
\end{eqnarray}
According to the $SU(3)_c$ and $SU(2)_s$ symmetry, the $SU(6)_{cs}$
eigenstate of the $q_2 \bar{q}_3 \otimes q_1 \bar{q}_4$ form is the
same as that of the $q_1 \bar{q}_3 \otimes q_2 \bar{q}_4$ form in
the first tetraquark state, while there appears an extra minus sign
in the second tetraquark state.


\begin{thebibliography}{10}



 \bibitem{Choi:2003}
S.K. Choi {\it et~al.}, Belle Collaboration,
 \newblock Phys. Rev. Lett. {\bf 91}, 262001 (2003).

\bibitem{R.Aaij:2013}
R.Aaij {\it et~al.}, LHCb Collaboration,
\newblock Phys. Rev. Lett. {\bf 110}, 222001 (2013).

 \bibitem{C.Y.Wong:2004}
C.Y.~Wong,
 \newblock Phys. Rev. {\bf C69}, 055202 (2004).

 \bibitem{E.S.Swanson:2004}
E.S.Swanson,
 \newblock Phys. Lett. {\bf B588}, 189 (2004).

 \bibitem{M.Suzuki:2005}
M.Suzuki,
 \newblock Phys. Rev. {\bf D72}, 114013 (2005).

 \bibitem{S.L.Zhu:2008}
S.L.Zhu,
 \newblock Int. J. Mod. Phys. {\bf E17}, 283 (2008).

\bibitem{B. Aubert:2005}
B. Aubert {\it et~al.}, $BARBAR$ Collaboration,
\newblock Phys. Rev. Lett. {\bf 95}, 142001 (2005).

\bibitem{C.Z. Yuan:2007}
C.Z. Yuan {\it et~al.}, Belle Collaboration,
\newblock Phys. Rev. Lett. {\bf 99}, 182001 (2007).

\bibitem{Z.Q. Liu:2013}
Z.Q. Liu {\it et~al.}, Belle Collaboration,
\newblock Phys. Rev. Lett. {\bf 111}, 019901 (2013).

\bibitem{B. Aubert:2007}
B. Aubert {\it et~al.}, $BARBAR$ Collaboration,
\newblock Phys. Rev. Lett. {\bf 98}, 212001 (2007).

\bibitem{X.L. Wang:2007}
X.L. Wang {\it et~al.}, Belle Collaboration,
\newblock Phys. Rev. Lett. {\bf 99}, 142002 (2007).

\bibitem{J.P. Lees:2014}
J.P. Lees {\it et~al.}, $BARBAR$ Collaboration,
\newblock Phys. Rev. {\bf D89}, 111103 (2014).

\bibitem{G.Pakhlova:2008}
G.Pakhlova {\it et~al.}, Belle Collaboration,
\newblock Phys. Rev. Lett. {\bf 101}, 172001 (2008).

 \bibitem{Ablikim:2013}
M.~Ablikim  {\it et~al.}, BESIII Collaboration,
 \newblock Phys. Rev. Lett. {\bf 110}, 252001 (2013).

 \bibitem{Z.Q. Liu:2013-2}
Z.Q. Liu {\it et~al.}, Belle Collaboration,
\newblock Phys. Rev. Lett. {\bf 110}, 252002 (2013).

\bibitem{T.Xiao:2013}
T. Xiao, S. Dobbs, A. Tomaradze and Kamal K. Seth,
 \newblock Phys. Lett. {\bf B727}, 366 (2013).

 \bibitem{Ablikim:2014}
M.~Ablikim  {\it et~al.}, BESIII Collaboration,
 \newblock Phys. Rev. Lett. {\bf 112}, 132001 (2014).

 \bibitem{M. Ablikim:2013}
M.~Ablikim  {\it et~al.}, BESIII Collaboration,
 \newblock Phys. Rev. Lett. {\bf 111}, 242001 (2013).

\bibitem{I.Adachi:2011}
I.~Adachi  {\it et~al.}, Belle Collaboration,
 \newblock arXiv:1105.4583 [hep-ex]

 \bibitem{R.Mizuk:2008}
R.~Mizuk {\it et~al.}, Belle Collaboration,
 \newblock Phys. Rev. {\bf D78}, 072004 (2008).

\bibitem{Choi:2008}
S.K. Choi  {\it et~al.}, Belle Collaboration,
 \newblock Phys. Rev. Lett. {\bf 100}, 142001 (2008).

 \bibitem{K.Chilikin:2013}
K.Chilikin  {\it et~al.}, Belle Collaboration,
 \newblock Phys. Rev. {\bf D88}, 074026 (2013).

\bibitem{zc4200}
G. Mohanty, \textit{Recent results on hot topics from Belle}, talk
in the Fifth Workshop on Theory, Phenomenology and Experiments in
Flavour Physics, May 23-25, Capri Island, Italy.

\bibitem{ZHUPLB2005}S. L. Zhu, Phys.Lett. B625, 212 (2005).

\bibitem{F.E.Close:2004}
F.E.~Close and P.R.~Page,
 \newblock Phys. Lett. {\bf B578}, 119 (2004).

 \bibitem{M.B.Voloshin:2004}
M.B.~Voloshin,
 \newblock Phys. Lett. {\bf B579}, 316 (2004).

 \bibitem{N.A.Tornqvist:2004}
N.A.T\"{o}rnqvist,
 \newblock Phys. Lett. {\bf B590}, 209 (2004).

 \bibitem{Y.R.Liu:2010}
Y.-R.Liu, M. Oka, M. Takizawa, X.Liu, W.-Z. Deng and S.-L. Zhu,
 \newblock Phys. Rev. {\bf D82}, 014011 (2010).

\bibitem{lining}
Ning Li, Shi-Lin Zhu,
\newblock Phys.Rev. {\bf D86}, 074022 (2012).

\bibitem{H.Hogaasen:2006}
H. Hogaasen, J.M. Richard and P. Sorba,
 \newblock Phys. Rev. {\bf D73}, 054013 (2006).

 \bibitem{D.Ebert:2006}
D. Ebert, R.N. Faustov and V.O. Galkin,
 \newblock Phys. Lett. {\bf B634}, 214 (2006).

 \bibitem{N.Barnea:2006}
N. Barnea, J. Vijande and A. Valcarce,
 \newblock Phys. Rev. {\bf D73}, 054004 (2006).

 \bibitem{Y.Cui:2007}
Y. Cui, X.L. Chen, W.Z. Deng and S.L. Zhu,
 \newblock High Energy Phys. Nucl. Phys. {\bf 31}, 7 (2007).

 \bibitem{R.D.Matheus:2007}
R.D. Matheus, S. Narison, M. Nielsen and J.M. Richard,
 \newblock Phys. Rev. {\bf D75}, 014005 (2007).

 \bibitem{T.W.Chiu:2007}
T.W.Chiu and T.H. Hsieh,
 \newblock Phys. Lett. {\bf B646}, 95 (2007).

 \bibitem{Q.Wang:2013}
Q. Wang, C. Hanhart and Q. Zhao,
 \newblock arXiv:1303.6355 [hep-ph]

  \bibitem{Oset}
F. Aceti, M. Bayar, E. Oset, A. Martinez Torres, K. P. Khemchandani,
F. S. Navarra and M. Nielsen,
 \newblock arXiv:1401.8216 [hep-ph]

\bibitem{zhaolu}
L. Zhao, L. Ma, and S.-L. Zhu,
\newblock Phys.Rev. {\bf D89}, 094026 (2014).

\bibitem{chenwei1}
W. Chen, T.G. Steele, M.-L. Du, and S.-L. Zhu,
\newblock Eur.Phys.J. {\bf C74}, 2773 (2014).

\bibitem{hejun}
J. He, X. Liu, Z.-F. Sun, and S.-L. Zhu,
\newblock Eur.Phys.J. {\bf C73}, 2635 (2013).


\bibitem{sun}
Z.-F. Sun, Z.-G. Luo, J. He, X. Liu, and S.-L. Zhu,
\newblock Chin.Phys. {\bf C36}, 194 (2012).

 \bibitem{C.D. Deng:2014}
C.D. Deng:2014, J.L. Ping and F. Wang,
 \newblock arXiv:1402.0777 [hep-ph]

  \bibitem{J.M. Dias:2014}
J. M. Dias, F. S. Navarra, M. Nielsen and C. Zanetti,
 \newblock arXiv:1311.7591s [hep-ph]

\bibitem{Y.R.Liu:2008}
Y.-R.Liu, X.Liu, W.-Z. Deng and S.-L. Zhu,
 \newblock Eur. Phys. J. {\bf C56}, 63 (2008).

 \bibitem{X.Liu:2009}
X.Liu, L.-Z. Gang, Y.-R.Liu, and S.-L. Zhu,
 \newblock Eur. Phys. J. {\bf C61}, 411 (2009).

 \bibitem{A.E.Bondar:2011}
A.E. Bondar, A. Garmash, A.I. Milstein, R. Mizuk and M.B. Voloshin,
 \newblock arXiv:1105.4437 [hep-ph]

\bibitem{liuxiang1}
Z.-F. Sun, J. He, X. Liu, Z.-G. Luo, and S.-L. Zhu,
\newblock Phys.Rev. {\bf D84}, 054002 (2011).

\bibitem{mali}
L. Ma, X.-H. Liu, X. Liu, and S.-L. Zhu,
\newblock arXiv:1404.3450 [hep-ph].

\bibitem{liuxiaohai}
X.-H. Liu, L. Ma, L.-P. Sun, X. L., and S.-L. Zhu,
\newblock arXiv:1407.3684 [hep-ph].

 \bibitem{R.L. Jaffe:1977-1}
 R.L. Jaffe,
 \newblock Phys. Rev. {\bf D15}, 267 (1977).

 \bibitem{R.L. Jaffe:1977-2}
 R.L. Jaffe,
 \newblock Phys. Rev. {\bf D15}, 281 (1977).

 \bibitem{K.T. Chao:1980}
 K.T. Chao,
 \newblock Nucl. Phys. {\bf B169}, 281 (1980).

 \bibitem{K.T. Chao:1981}
 K.T. Chao,
 \newblock Nucl. Phys. {\bf B183}, 435 (1981).

\bibitem{Y. Cui:2006}
 Y. Cui, X. L. Chen, W. Z. Deng and S. L. Zhu,
 \newblock Phys. Rev. {\bf D73}, 014018 (2006).

\bibitem{R.L. Jaffe:1977-3}
R.L. Jaffe,
 \newblock Phys. Rev. Lett. {\bf 38}, 195 (1977).

\bibitem{T. Degrand:1975}
 T. Degrand, R.L. Jaffe, K. Johnson and J. Kiskis
 \newblock Phys. Rev. {\bf D12}, 2060 (1976).

 \bibitem{D. Strottman:1976}
 D. Strottman,
 \newblock J. Math. Phys. {\bf 20}, 1643 (1976).

 \bibitem{S. I. So:1976}
 S. I. So, D. Strottman,
 \newblock J. Math. Phys. {\bf 20}, 153 (1976).

  \bibitem{PDG}
K. Nakamura, {\it et~al.}, Particle Data Group,
 \newblock J. Phys. {\bf G37}, 075021 (2010).

\bibitem{chenwei}
W. Chen, S.-L. Zhu,
\newblock Phys.Rev. {\bf D83}, 034010 (2011).










\end{thebibliography}
\end{document}